# Structural, Raman and photoluminescence studies on nanocrystalline diamond films: Effects of ammonia in feedstock

K. Ganesan,[a,*] P. K. Ajikumar,[a] S.K.Srivatava,[b] and P. Magudapathy [b]

[a] *Surface and Nanoscience Division, Materials Science Group, Indira Gandhi Centre for Atomic Research, HBNI, Kalpakkam - 603102, India.*

[b] *Materials Physics Division, Materials Science Group, Indira Gandhi Centre for Atomic Research, Kalpakkam - 603102, India.*

**Abstract**

Herein, we report on the improvement of structural quality and enhancement of photoluminescence (PL) emission for an optimally *N* doped nanocrystalline diamond (NCD) film. Pure and *N* doped nanocrystalline diamond films are synthesized on Si by hot filament chemical vapour deposition using $NH_3:CH_4:H_2$ at different nominal N/C ratios viz. 0, 0.13, 0.35, 0.50 and 0.75 in feedstock. X-ray diffraction analysis reveal a systematic initial increase and then a decrease in crystallite size with N/C ratio in feedstock. Further, a monotonic increase in Raman line width and peak position of diamond band indicates that the compressive strain in diamond lattice increases as a function of N/C ratio upto 0.50. However, at higher N/C ratio of 0.75, the compressive strain gets relaxed a little and produces a lower strain. Furthermore, a unique Raman mode at 1195 $cm^{-1}$ is observed corresponding to the C=N-H vibrations indicating a significant *N* concentration in the NCD films. In addition, visible and UV PL studies reveal the presence of several *N*-related colour centres with multiple emission lines in the range of 380 – 700 nm. An optimally *N* doped diamond film grown at N/C ratio of 0.35 in feedstock shows a significant enhancement in room temperature PL emission at ~ 505 and 700 nm. This PL enhancement is attributed to H3 and other aggregates of *N* related defect centres, under 355 and 532 nm laser excitations respectively.

---

[*] Corresponding author. Email : kganesan@igcar.gov.in (K.Ganesan)

# 1. Introduction

Diamonds have attracted much attention of researchers due to their unique structural, electrical, optical, chemical and mechanical properties. The single crystal diamonds are used in several electronic devices with high performance which can operate at high temperature, high frequency and high power and high energy ionizing radiation. Moreover, it has applications in particle detectors, single photon emitters in quantum information technology [1,2,3]. Whereas the nanocrystalline diamonds (NCDs) find applications in field emission, tribology, passive element as heat spreading in electronic devices and optical coatings [4,5]. Even though diamond has several unmatched properties in comparison to its competing materials, still a lot challenges have to overcome in terms of growth and doping to use it in practical device applications. Over the past few decades, the research and technology on diamond materials upsurge rapidly since the ability to produce synthetic diamonds with very high purity and structural quality that are almost similar to natural diamond.

Hot filament chemical vapor deposition (HFCVD) is one of the simplest techniques and it can produce diamond films with reasonable high quality and growth rate. Boron doping in diamond is the most successful method for p-type conduction with its ionization energy is ~ 0.37 eV. Mostly *N* is the preferred impurity for *n*-type doping, however, *N* does not provide excess carrier at room temperature due to its deep energy level (1.7 eV) in diamond. Hence, *N* doped diamond is not so beneficial for electronic device applications. However, *N* doping in NCD and ultra nanocrystalline diamond films enhance the electrical conductivity by several orders and also behaves as n-type semiconductors. The enhanced conductivity is generally assigned to increase in $sp^2$–rich C=C bonding at grain boundaries (GBs) [6]. Further, N doping in NCDs enhances physical and mechanical properties which make them useful for applications in photocathode for photoinjectors [7], electrochemical devices that can work under higher potential window [8], micro-electromechanical system [9], and biomedical devices [5]. Due to these several advantages of *N* doped NCD, the research interest on the study of *N* incorporation and its effects on growth, structural, electrical and optical properties of diamond is ever expanding. A brief glimpse of the existing literature related to the effects of *N* in the feed gas on the growth and properties of diamond are summarized here.

Generally $N_2$ or $NH_3$ is used as source of *N* under plasma assisted CVD reactors. $NH_3$ is preferred over $N_2$ because of weaker N-H bond as compared to stronger N≡N bond of $N_2$ under HFCVD reactor [10]. The reported N/C ratio in the feedstock is varied from as small as 0.1 to greater than 150 % for the growth of *N* doped diamonds [10-21]. The presence of *N* in feedstock significantly affect the growth mechanism, morphology, growth rate, grain size, chemical bonding at grain and GBs. It is mostly noticed that a small amount of *N* in feed gas increases the growth rate remarkably and also the growth rate decreases at higher N/C ratio in feedstock [11-14]. Further, the morphology and growth habit change due to the orientation dependent growth rate which results in texturing of diamond films [11,12,13,14,15,16]. Also, the presence of *N* in the feedstock increases the amount of H impurity incorporation in diamond during growth [17]. Further, *N* incorporation in feedstock is found to either improve or deteriorate the structural quality of diamond upon *N/C* ratio and other feedstock composition [10,18]. Moreover, diamond films are also grown under *N* rich environments (N/C ratio > 100 % in feedstock) which results in nanocrystalline structure with high conductivity [16,19]. Furthermore, the grain size is noticed to decrease significantly with a small amount of *N/C* ratio in the feedstock [17]. A few study reports the increase in grain size with *N/C* ratio in feedstock [14,19].

Intensive vibrational spectroscopic studies are carried out to assess the structural quality of the diamonds grown with different *N* concentration in the feedstock using visible and UV Raman spectroscopy and IR spectroscopy. Nitrogen incorporation in diamond lattice results in significant mechanical stress, increase in vacancy defects and also N-vacancy defect complexes. Consequently, *N* doping in diamond decreases the phonon life time which manifest as broadening of Raman line width [20]. In addition, a characteristic vibration at 1190 $cm^{-1}$ is observed for *N* doped diamond that arises due to C=N-H bonding at grain and GBs [10,21]. Also, the vacancies and N-related defects create deep energy levels within the forbidden energy gap and each such defect (known as color centers) emits photons at characteristic wavelength under suitable excitation. Thus, many investigations report on the identification of *N* induced point defects and N-related complex defects in diamonds through optical spectroscopic studies [22,23,24,25,26,27,28,29,30]. The most common defects observed in *N* doped diamonds are neutral & negatively charged vacancies ($V^0$ & $V^-$), N-Vacancy complex defects ($NV^0$ & $NV^-$), H3 (two substitutional *N* atoms separated by a vacancy, N-V-N), H4 (four substitutional *N* atoms surrounding two lattice vacancies, 4N+2V), N2 center (neutral nearest neighbor of two substitutional *N* atoms), and N3 (three substitutional *N* atoms surrounding a vacancy, 3N+V).

Although a significant amount of literature is currently available on the role of *N* containing molecules in feedstock on the growth and properties of *N* doped NCDs, a combined study on structural and optical properties of N-doped NCDs is limited. In this study, *N* doped NCD films are synthesized by HFCVD using wide range of N/C ratios upto 75% in the feedstock. The *N* induced structural evolution is studied systematically using X-ray diffraction (XRD), scanning electron microscopy (SEM), and Raman spectroscopy. Further, *N* related impurity energy levels and their influence on the photoluminescence (PL) intensity are measured through PL spectroscopy using 532 and 355 nm diode lasers. Further, a combined analysis of XRD, SEM, Raman and PL are performed to have better understating on the role of *N* in feedstock on the growth, structure and optical properties of *N* doped NCD films.

## 2. Experimental methods

Nitrogen doped diamond films were grown by a custom designed HFCVD system and the details could be found elsewhere [31,32]. The diamond films are grown on Si (111) and the substrates were chemo-mechanically polished with micron level diamond paste prior to growth. The undoped and N doped NCD films were grown by admitting feedstock gases of $CH_4$, $H_2$ and $NH_3$ at an appropriate concentrations as per the details given in Table 1. The operating pressure was maintained at 30 mbar and the substrate temperature was controlled to be constant at 800 $^0$C using an additional resistive heater throughout the synthesis. An identical growth condition was maintained for all growth experiments with a fresh filament, the carburization condition, filament power, filament to substrate distance and growth time in order to avoid the variation due to instrumental parameters. The growth experiments were carried out for 6 hours. A uniform diamond films with very good adhesion was obtained over the entire area of the substrates. The morphological and microstructural properties of the grown diamond films were studied by SEM (Supra 55, Carl Zeiss, Germany) and XRD ( Inel, equinox 2000 ). Raman and PL spectroscopic studies of these diamonds were carried out using micro-Raman spectrometers (M/s Witech spectrometer, Germany and M/S Invia, Renishaw, UK) using 355 and 532 nm diode lasers.

Table 1. Growth parameters for pure and N doped nanocrystalline diamond films

| Sample name | Actual feedstock flow rate (sccm) | | | N/C ratio (%) | C:H:N ratio | Crystallite size (nm) | Thickness (μm) |
|---|---|---|---|---|---|---|---|
| | $CH_4$ | $H_2$ | $NH_3$ | | | | |
| N0  | 2 | 100 | 0    | 0    | 2:204:0       | 14.2 | 2.2 |
| N13 | 2 | 100 | 0.25 | 12.5 | 2:204.75:0.25 | 14.8 | 2.7 |
| N35 | 2 | 100 | 0.70 | 35.0 | 2:206.1:0.70  | 16.3 | 1.4 |
| N50 | 2 | 100 | 1.00 | 50.0 | 2:207.0:1.00  | 12.0 | 1.1 |
| N75 | 2 | 100 | 1.50 | 75.0 | 2:208.5:1.50  | 11.3 | 1.4 |

## 3. Results and discussion

### 3.1. X-ray diffraction

Fig. 1 depicts the XRD pattern of the diamond films grown under different N/C ratios. These XRD spectra exhibit a typical pattern of polycrystalline cubic diamond with diffraction peaks at ~ 43.6, 74.9 and 91.0 degrees corresponding to (111), (220) and (311) planes, respectively. Further, the full width at half maximum (FWHM) of the (111) diffraction peak varies as a function of N/C ratio as shown in inset of Fig.1. The FWHM of N35 film is the lowest and it increases on either lower or higher N/C ratios in the feedstock. The average crystallite size is calculated using Scherrer formula and tabulated in Table 1. As shown in Table 1, the crystallite size increases initially upto N/C ratio of 0.35 and then, it decreases for higher N/C ratios. Further, a large amorphous background in the diffraction pattern is also observed for the diamond film grown without any $NH_3$ in feedstock. However, the amorphous background signature decreases significantly by addition of $NH_3$ in feedstock, as can be seen from Fig. 1. Note that the X-rays may penetrate to the defective nucleation layer at film – substrate interface that can affect the XRD patterns since the thicknesses of the NCD films are in the range of 1.1 – 2.7 μm. However, the thicker film, N0, shows a higher amorphous like background signature than the thinner films such as N35, N50 & N75. Hence, we conclude that the observed XRD patterns mainly arising from bulk of the films rather than the defective interface layer. Consequently, the observed variation in amorphous signature is attributed to the nature of chemical bonding in NCD films grown with and without addition of $NH_3$ in feedstock. These aspects are further discussed in the forthcoming section under Raman spectroscopy.

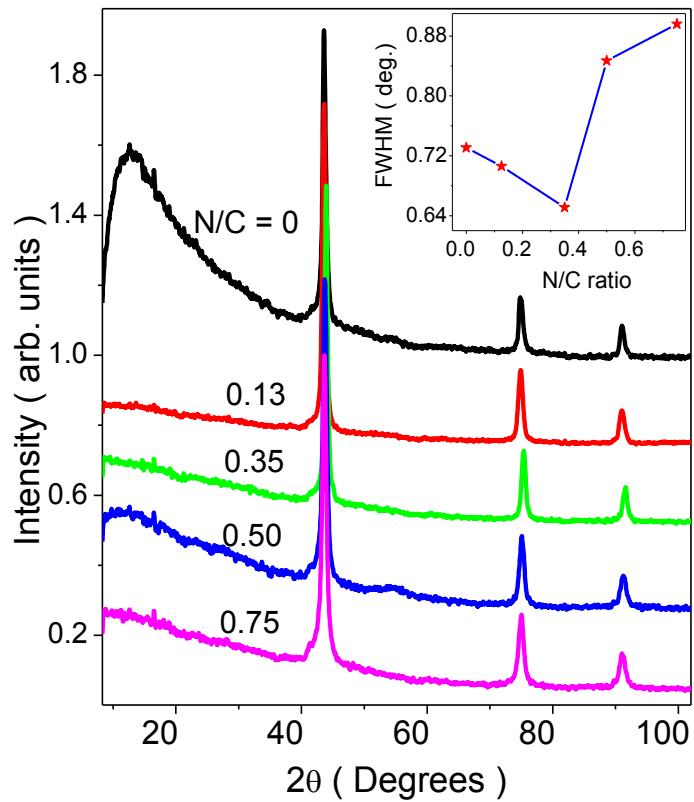

Fig.1. XRD patterns of the diamond films grown under different N/C ratios. The inset shows the variation of full width at half maximum of (111) peak as a function of N/C ratio. These spectra are normalized with (111) intensity and made a vertical shift for clarity.

### 3.2. Scanning electron microscopy

Figure 2a-2e shows the SEM micrographs of the grown diamond films at different N/C ratios in the feedstock. The observed microstructure of these films is found to be nano-crystalline nature, in general. However, it is difficult to measure the actual grain size in these NCD films because of the limitations of SEM to distinguish grain and GBs. The variation of film thickness as a function of N/C ratio in the feedstock is shown in Fig.2f. At N/C ratio of 0.13, the growth rate is higher than the diamond film that is grown without $NH_3$ in the feedstock. On the other hand, the growth rate decreases significantly when the N/C ratio is more than 0.13 in the feedstock (Fig. 2f). In CVD, the growth rate is determined

by the competing mechanisms of simultaneous growth and etching processes that occur on the growing surface. A small addition of ammonia in feedstock increases the N containing radicals such as CN and HCN which enhance the abstraction of H from growing surface by forming HCN and H2CN radicals. The process of enhanced H abstraction by N containing radicals increases the surface adsorption of $CH_3$ radicals which results in the increase of growth rate [10,13,14,33]. On the other hand, at higher N/C ratio of > 0.13 in feedstock, the formation of HCN radicals significantly increases and subsequently the amount of $CH_x$ radicals in the reactor decreases considerably which results in reduction of growth rate. In addition, the *N* radicals also promote graphitic C=N bonding that prevents further diamond growth and also allows the formation of secondary nucleation. These factors decrease the growth rate as well as tune the morphology at higher N/C ratios in the feedstock.

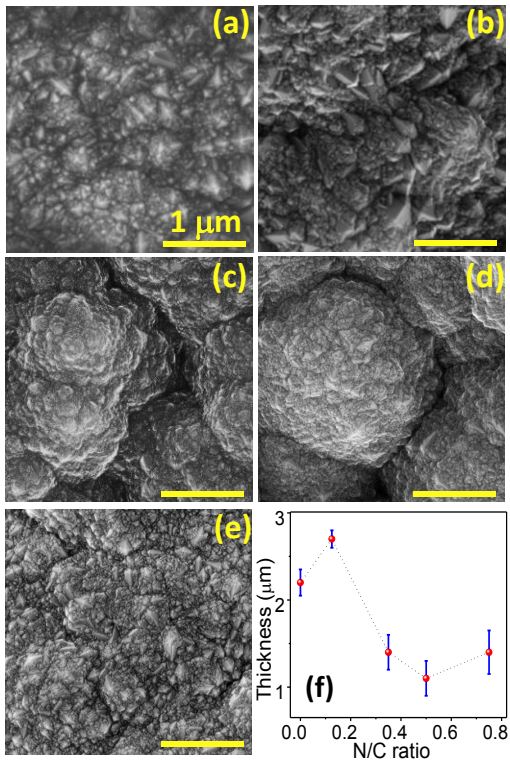

Fig. 2. SEM micrographs of nanocrystalline diamond films grown under different N/C ratios in the feedstock at (a) 0, (b) 0.13, (c) 0.35, (d) 0.50 and (e) 0.75. (f) The variation of thickness of the diamond films versus N/C ratio in the feedstock during growth. The scale bar is same for all the SEM micrographs.

### 3.3. Raman spectroscopy

Figure 3a and 3b represent the Raman spectra of the diamond films recorded using 532 and 355 nm wavelength lasers, respectively. The visible wavelength Raman spectra (Fig 3a) display a typical characteristics of NCD structure with a weak diamond Raman band ~ 1332 cm$^{-1}$ embedded with strong and broad D and G bands at ~ 1350 and 1580 cm$^{-1}$ respectively, of disordered graphitic carbon. In addition, the signature of trans-polyacetylene (TPA) structure is also evidenced from the Raman bands of $\nu_1$ and $\nu_3$ at 1140 and 1480 cm$^{-1}$ respectively, in these spectra. The evolution of TPA structure mainly depends on the amount of H in the disordered carbon network especially at GBs [34]. As can be evidenced from the Fig. 3a, the signature of TPA structure is negligible for N35 film while it is moderate for N13 and N50 and larger for other diamond films, N0 and N75. The observed behavior of TPA bands also corroborate well with the microstructure analysis in terms of grain size as measured by XRD. Moreover, an additional weak Raman band appears at about 1190 cm$^{-1}$ for N35 and N50 films, as indicated by * mark in Fig 3a. This band arises due to the characteristic signature of C=N-H vibration in the diamond grain and GBs [10,21]. The necessary condition for the appearance of this band is correlated with the presence of C-N bonding with increase in graphitic phase and also (100) orientation in the films [10]. On the other hand, the films N13 and N75 have larger amount of TPA which signifies a higher C-H bonding with sp$^3$ hybridization in N13 and N75 as compared to N35 and N50 films. Consequently, the 1190 cm$^{-1}$ band is not observed in N13 and N75 films. Obviously, the band at 1190 cm$^{-1}$ is not observed for N0 film because of zero ammonia in feedstock.

Further, the Raman spectra also consist of a weak Raman signal from Si substrate indicating that the penetration depth of laser is higher than the thickness of the studied NCD films. Under such condition, the laser also probes the defective nuclear later at film – substrate interface which can influence the overall Raman spectra of the NCD films. However, since our samples are nanocrystalline in size with a large amount of sp$^2$-rich grains and grain boundaries in the bulk which can absorb laser intensity significantly, the contribution from defective nucleation layer, if at all present, is negligible by considering the volume ratio of bulk to interface layer [35]. Hence, it is understood that the Raman spectra of the studied diamond films are not significantly affected by film thickness; rather the spectra mainly arise by the nature of C-C bonding in the bulk of the films. Since Raman scattering cross section of sp$^2$-rich carbon bonds is much higher ( upto 230 times) than that of $sp^3$ type C-C bonds, graphitic structure dominates over diamond structure in visible laser Raman spectra of the NCD films. Hence, UV Raman spectroscopy is also performed for these NCD films and discussed below.

Fig.3b displays micro-UV Raman spectra of diamond films grown under different N/C ratios in feedstock which shows a prominent Raman band at ~ 1332 cm$^{-1}$ along with disordered $sp^2$-rich graphitic carbon bands. Further, the Raman spectra are deconvoluted into five bands and the best fit parameters are calculated and given in Table 2. The observed Raman peaks are centered at about 1200, 1332, 1350, 1550 and 1580 cm$^{-1}$ for all the N-doped diamond films. The bands at ~ 1200 and 1550 cm$^{-1}$ are assigned to $\nu_1$ and $\nu_3$, respectively of the TPA structure and they are upshifted due to the dispersion nature of these bands with excitation energy [34]. The other two bands at ~ 1350 and 1580 cm$^{-1}$ are due to D and G bands of disordered graphitic carbon. Further, the intensity ratio between D and G ($I_D/I_G$) is commonly used to quantify the disorder in graphic system. Since the visible Raman spectra are completely dominated by disorder, it is difficult to deconvolute the spectrum and hence, UV-Raman spectra are used to calculate the $I_D/I_G$ ratio and the results are shown in Table 2. Here, the $I_D/I_G$ ratio varies from 0.81 to 1.42 depending upon N concentration in the feedstock. This variation clearly indicates the differences in the graphitic content in diamond films. The $I_D/I_G$ ratio for the film N75 is 0.81 while it is 1.42 for N50 film. Thus, N75 has lower graphitic content than the N50 and other diamond films. In addition, the N35 and N50 films have relatively lower FWHM of G band and higher $I_D/I_G$ ratio which indicate a higher graphitic content in the films [36]. This observation is also consistent with the appearance of 1190 cm$^{-1}$ peak under 532 nm laser excitation in these films because C=N-H vibration is only evident when the diamond matrix has increased amount of graphitic phase and (100) orientation [10].

The Raman line width and peak position of the *N* doped diamonds exhibit a monotonic increase with *N* concentration upto N/C ratio of 0.50, as can be seen from Fig.3c and Table 2. The increase in line width indicates the increase in disorder in NCD lattice with nominal N/C ratio in the feedstock. Also, the lattice strain increases the phonon scattering probability and thus, the phonon life time decreases that results in broadening of Raman band [20]. Such a linear increase in line width with *N* concentration has been well documented for the *N* doped single crystal diamonds but a similar study on polycrystalline diamonds is scarce and also the analysis becomes complex due to other structural defects [37,38]. The present study also confirms a similar trend in broadening of Raman line width for *N* doped NCD films. However, Raman peak position of these diamond films exhibit a blue shift and it is in contrast to existing literature on single crystal diamonds which shows either no change in peak position for relatively low N concentration [20,37] or a red shift of ~ 0.4 cm$^{-1}$ for higher N concentration [38]. The observed blue shift in Raman peak position indicates that the diamond lattice undergoes a compressive strain with *N* doping in diamond. At higher N/C ratio of 0.75 in feedstock, both line width and peak position show a little reverse trend indicating the higher phonon life time and lower compressive strain as compared to N50 film.

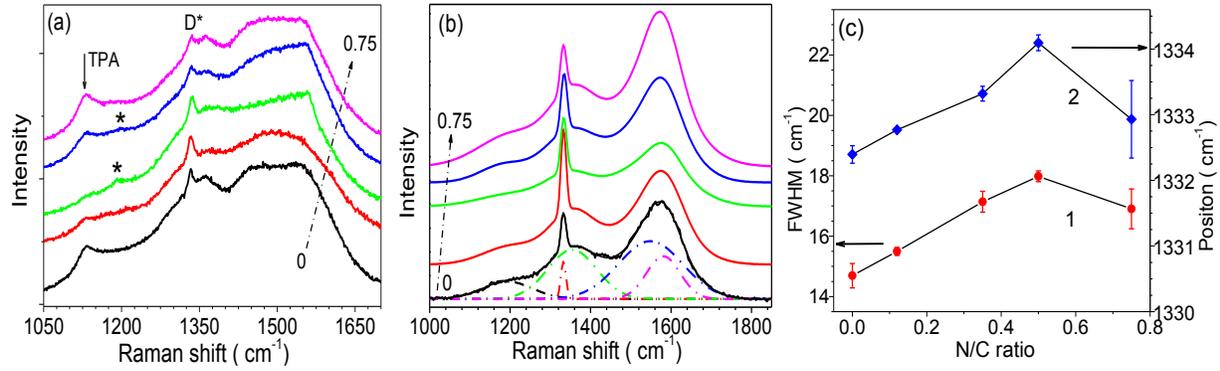

Fig.3. Raman spectra of diamond films recorded with laser wavelength of (a) 532 nm and (b) 355 nm; (c) The variation of full width at half maximum and peak position as a function of N/C ratio in feedstock for the diamonds grown on Si, calculated from 355 nm excitation. The dotted line profiles in Fig.3b are the deconvolution curves with Gaussian function to the experimental data. The Raman spectra are recorded under identical conditions and shifted vertically for clarity. The solid lines in Fig.3c are only guideline to eye.

Table 2. UV micro-Raman fit parameters for the diamond films grown under different N/C in feedstock

| Sample | $\nu_1$ band (cm$^{-1}$) | | D* (cm$^{-1}$) | | D band (cm$^{-1}$) | | $\nu_3$ (cm$^{-1}$) | | G band (cm$^{-1}$) | | $I_D/I_G$ ratio |
|---|---|---|---|---|---|---|---|---|---|---|---|
| | Pos. | $\Gamma$ | Pos. | $\Gamma$ | Pos. | $\Gamma$ | Pos. | $\Gamma$ | Pos. | $\Gamma$ | |
| N0 | 1193.8 | 123.8 | 1332.5 | 12.7 | 1354.3 | 122.1 | 1549.0 | 159.6 | 1580.8 | 83.9 | 1.15 |
| N13 | 1206.0 | 142.3 | 1332.8 | 13.1 | 1354.5 | 111.3 | 1546.7 | 172.5 | 1582.9 | 87.6 | 1.10 |
| N35 | 1215.3 | 176.5 | 1333.2 | 14.3 | 1348.4 | 114.2 | 1545.9 | 184.4 | 1585.3 | 75.7 | 1.23 |
| N50 | 1184.8 | 118.3 | 1334.0 | 15.4 | 1352.6 | 132.0 | 1555.1 | 156.9 | 1581.6 | 75.7 | 1.42 |
| N75 | 1211.3 | 170.8 | 1332.2 | 14.9 | 1355.7 | 112.4 | 1542.1 | 188.4 | 1578.3 | 84.8 | 0.81 |

The Raman line broadening and position shift in NCD films can be associated with several factors such as surface and bulk structural defects, GBs, internal strain caused by impurities, growth temperature induced thermal stress that arises due to the difference in thermal expansion coefficients (TECs) of substrate and diamond and also laser induced heating on the surface [38,39]. However, the analysis of peak shift and broadening of diamond Raman band in NCD films is complex. In the present study, the laser induced temperature effect can be ignored since it always redshifts the Raman peak position which is about 0.5 cm$^{-1}$ for an increase in temperature upto 500 K [38]. The compressive stress can be attributed to the nature of chemical bonding in the grains and GBs that consist of highly cross-linked sp$^2$ phase network surrounding the nanodiamond particles [35]. Further, the compressive stress is also directly proportional to the $I_D/I_G$ ratio, which signifies the amount of sp$^2$ bonding in the films, as can be evidenced from Table 2. Moreover, the biaxial thermal stress due to TEC mismatch between Si and NCD film can also induce compressive stress but it is again linked with the presence of sp$^2$ phases through Young modulus of the NCD films which can vary from 200 to 800 GPa [39]. However, a reduction in line width with increase in doping concentration (N/C ratio of 0.75) is possible when the lattice gets relaxed by defect induced stabilization of diamond [11]. Indeed, a large amount of point defects are also created in diamond lattice during growth. The addition of impurities in feedstock can minimize such point defects by forming impurity – vacancy pairs in the lattice. To verify this fact further, visible and UV PL measurements are performed because PL emission spectra provide a valuable information about the population density of *N* states within the forbidden band gap and the results are discussed in the forthcoming section.

### 3.4 Photoluminescence spectroscopy

Figure 4a shows the room temperature PL spectra of the diamond films grown at different N/C ratio in feedstock by an excitation wavelength of 532 nm. The undoped diamond film N0 displays two broad PL bands near 630 and 700 nm which are typical for NCD structure [32]. The PL intensity enhances significantly in the studied wavelength range at a small amount of *N* doping ( N/C ratio = 0.13) in diamond lattice and also the maximum peak is shifted from about 630 nm to 700 nm. A further increase in *N* doping (N35), the PL intensity increases drastically with a peak at around 700 nm which is attributed to the aggregates of nitrogen defects such as N2, N3, N4 centers [23]. For even higher *N* doping (N50 and N75), the PL intensity of the 700 nm peak decreases monotonically with a change in line shape in the long wavelength regime. A closer look at the PL spectra clearly reveal the appearance of several bands near 630, 637, 660, 680 and 738 nm. The PL bands at 637 and 738 nm are assigned to zero phonon lines of $NV^-$ and negatively charged silicon – vacancy ($SiV^-$) color centers, respectively. The other bands at ~ 630, 660 and 680 nm are assigned to phonon side bands of $NV^0$ and $NV^-$ centers [24].

Fig. 4b shows the PL spectra of the N-doped diamond films excited by 355 nm. As similar to 532 nm excitation, the films N0 and N13 also exhibit a very broad PL emission in the wavelength range of 450 – 800 nm with a maximum at around 600 nm. However, PL intensity of N13 is much higher than that of N0 film. A further increase in *N* doping (N35), the PL intensity increases drastically and also exhibits a narrower peak at around 505 nm corresponding to H3 center, two substitution *N* atoms separated by a vacancy, N-V-N [25]. At even higher doping ( N50 & N75), the intensity of H3 emission decreases monotonically, which is a similar trend to the 700 nm emission band under 532 nm excitation, as can be seen from Fig.4b. As depicted in Fig. 4a and 4b, the reason for the reduction in the intensity is correlated with the increase in *N* - aggregated defects such as N2, N3, N4 centers that quenches the luminescent intensity of H3 centers [26,27]. Note that the increase in nominal N/C ratio in feedstock is known to increase the *N* concentration in diamond lattice [10]. However, the actual *N* concentration in diamond lattice is expected to be very low since the doping efficiency of *N* in diamond is very low (~$10^{-4}$) [11]. Furthermore, as shown in Fig.4a, the increase in 637 nm emission intensity as a function of *N* doping also reconfirms the increase in *N* concentration in the lattice. Thus, an optimal amount of *N* doping in diamond enhances PL emission in the wavelength region of 500 and 700 nm for the excitation of 355 and 532 nm, respectively.

A closer look at the Fig.4b and 4c indicate the presence of several color centers with PL emissions at different wavelengths ~ 389, 394-399, 412, 420, 450, 468, 480, and 495 nm. The peak at 389 nm is assigned to the pair formation of single substitutional *N* atom and an interstitial C atom in the nearest neighbor [28]. The emission lines at about 480 and 495 nm are assigned to N2 center (neutral nearest neighbor of two substitutional *N* atoms) and H4 center (four substitutional *N* atoms surrounding two lattice vacancies, 4N+2V) respectively [22,29]. The broad emission line ~ 420 nm is observed on all the diamond films irrespective of *N* concentration and it is attributed to the band-A emission which arises due to the $sp^2$ defects in the incoherent GBs and dislocations in the NCD films [30]. An another broad band in the range of ~ 394 – 400 nm can arise due to the combination of 394 nm emission (negatively charged vacancy center ($V^-$)), and N-band that occurs at ~ 400 nm due to the interstitial *N* atoms in diamond lattice [23]. A small hump at ~ 412 nm, as indicated by * mark in Fig.4c, can be associated with 389 nm sharp emission line and also it can be due to N3 center (three substitional *N* atoms surrounding a vacancy, 3N+V) [23]. The other bands at ~ 450 and 468 nm are associated with phonon side bands of N3 and N2 centers. Thus, PL spectra clearly indicate that the increase in N/C ratio in feedstock progressively increases the *N* concentration in the diamond lattice. Consequently, the point defects and N-related defect aggregates increase in diamond lattice with N/C ratio in feedstock. Also, these defects manifest as colour centers and exhibit an interesting luminescent characteristics which can be controlled by the amount of *N* concentration in the diamond lattice. Further, these results conclude that an optimal *N* doping in NCD improves the structural and optical emission characteristics significantly.

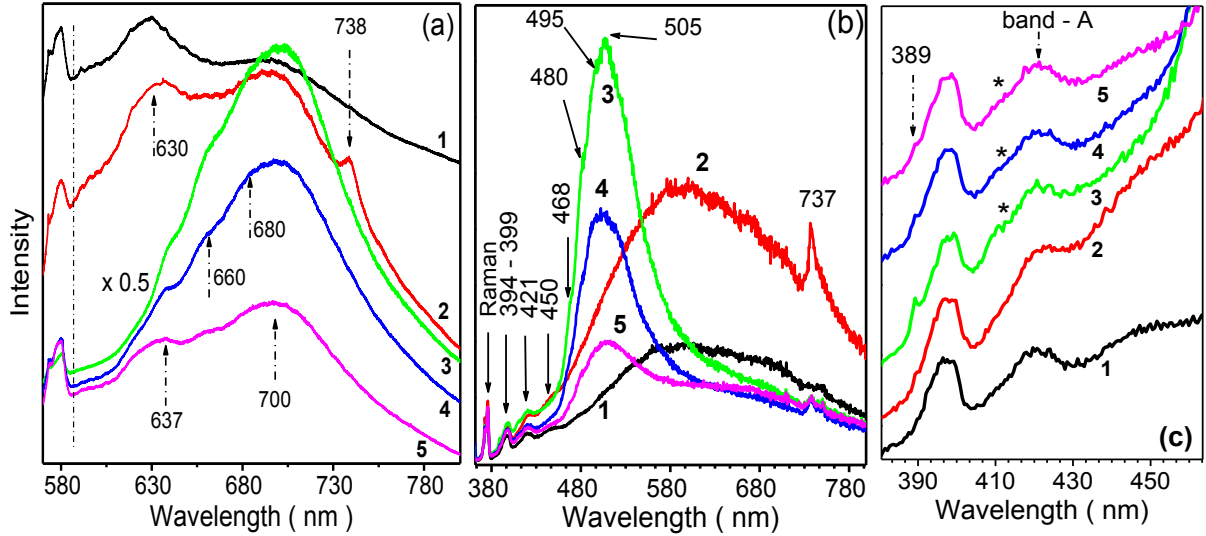

Fig.4. Photoluminescence spectra of *N* doped diamond films recorded under the excitation of (a) 532 and (b) 355 nm. (c) A magnified part of the PL spectra given in Fig.4b at lower wavelength. The numbers 1, 2, 3, 4 and 5 in the spectra represent the N/C ratio in the feedstock of 0, 0.13, 0.35, 0.50 and 0.75, respectively. All the spectra in each category are recorded at an identical condition. In Fig.4a, the PL intensity for N35 film is multiplied with 0.5 to bring down to scale and the spectra for N0 and N13 films are shifted vertically for clarity. The spectra in Fig.4c are shifted vertically for clarity. No background correction is performed on these spectra.

## 4. Conclusions

The pure and *N* doped nanocrystalline diamond (NCD) films are successfully synthesized by hot filament chemical vapour deposition using $NH_3:CH_4:H_2$ at different nominal N/C ratios in the feedstock. The structural quality of the *N* doped diamond films is found to improve at lower N/C ratio and it degrades at higher N/C ratio in feedstock as evidenced by x-ray diffraction. The N doped NCDs are under compressive strain due to the presence of large amount of $sp^2$-rich grains and grain boundaries and it manifests as increase in the Raman line width. In addition, the observation of a unique Raman mode at 1195 $cm^{-1}$ also confirms a significant *N* concentration in the NCD with C=N-H bonding. In addition, the *N* doped diamonds display several N-related colour centres with multiple emission lines in the range of 380 – 700 nm as evidenced by visible and UV photoluminescence (PL) spectroscopy. Especially, the intensity of PL emission at ~ 505 and 700 nm is increased drastically for an optimally *N* doped diamond film grown at N/C ratio of 0.35 in feedstock at room temperature under 355 and 532 nm laser excitations, respectively. Since the diamonds are known as radiation hard materials, an enhanced PL emission with improved structural quality of an optimally *N* doped diamonds have potential applications in luminescence based radiation detectors which can work under extreme environments.

**Declaration of competing interest**

There are no conflicts to declare.